# Power, Delay and Area Comparisons of Majority Voters relevant to TMR Architectures

P. BALASUBRAMANIAN*, N.E. MASTORAKIS[§¶]
* School of Computer Engineering
Nanyang Technological University
50 Nanyang Avenue
SINGAPORE 639798
Email: balasubramanian@ntu.edu.sg
[§] Department of Computer Science
Military Institutes of University Education
Hellenic Naval Academy
Piraeus 18539, GREECE
Email: mastor@hna.gr
[¶] Department of Industrial Engineering
Technical University of Sofia
Sofia 1000, Boulevard Kliment Ohridski 8
BULGARIA
Email: mastor@tu-sofia.bg

*Abstract:* - N-modular redundancy (NMR) is commonly used to enhance the fault tolerance of a circuit/system, when subject to a fault-inducing environment such as in space or military systems, where upsets due to radiation phenomena, temperature and/or other environmental conditions are anticipated. Triple Modular Redundancy (TMR), which is a 3-tuple version of NMR, is widely preferred for mission-control space, military, and aerospace, and safety-critical nuclear, power, medical, and industrial control and automation systems. The TMR scheme involves the two-times duplication of a simplex system hardware, with a majority voter ensuring correctness provided at least two out of three copies of the hardware remain operational. Thus the majority voter plays a pivotal role in ensuring the correct operation of the TMR scheme. In this paper, a number of standard-cell based majority voter designs relevant to TMR architectures are presented, and their power, delay and area parameters are estimated based on physical realization using a 32/28nm CMOS process.

## 1 Introduction

Radiation hardening by design at the component or module level is widely used to mitigate the problem of single event upsets (SEUs) and single event transients (SETs) in the case of ASICs and FPGAs. SETs, which occur due to high energy particle strikes, might cause a bit-flip at a gate output node or in interconnects formed between logic elements. A SET possessing sufficient amplitude and duration may be captured by a state-holding element in a system stage and subsequently latched, resulting in an error called as SEU. A SEU may also occur when a radiation phenomenon tends to directly flip the binary data output of a sequential element, which could immediately result in an error. SEUs tend to affect data processing in the successive system stage by allowing computation with erroneous data. With respect to radiation hardening by design, circuit/system level solutions exist for both ASICs and FPGAs.

In the case of ASICs, one common circuit level solution deals with the full-custom development of radiation-tolerant standard library cells which are meant for use in an ASIC-based synthesis environment. Companies such as Atmel [1] and BAE Systems [2] have developed radiation-tolerant standard cell libraries, which are available from 180nm down to the 45nm technology node for use in space and aerospace applications. However, radiation-tolerant cell designs tend to use extra transistors, adopt transistor sizing, and add extra capacitive loads to the output [3] – [5], and hence they are likely to occupy more area, consume more power and may be slower in comparison with commercial standard cell libraries besides being expensive. The other viable alternative for designers is to opt for a well-established circuit/system level solution such as TMR [6] [16], which necessitate triplication of a circuit or system and require a majority of them to maintain the correct operation.

Critics of TMR often point to the excess hardware overhead (about 200%) incurred. To minimize this hardware overhead, approaches for selective insertion of TMR have been proposed in the literature [7] – [9]. Selective TMR insertion entails the task of determining critical portions of a circuit/system where TMR can be applied to compensate for lesser fault masking, and non-critical portions where TMR need not be applied due to greater fault masking [10]. Although not all errors tend to get eliminated through selective TMR insertion, the overall error rate gets reduced [11]. Selective TMR insertion could be a feasible solution to alleviate the overheads of full TMR, especially for those applications where weight, cost, and performance also matter besides reliability and fault tolerance, such as medical, mobile, wearable and portable electronics, and electronics meant for military purposes.

For mission-critical systems, where reliability is paramount over cost, full TMR is highly preferred and has been chosen for many space and aerospace applications, from the design of the Saturn V Launch Vehicle Digital Computer [12] to the in-flight system design for the Mars Mission [13], and potentially even beyond. There are many practical situations where full TMR has been adopted for the development of radiation-hardened communication ICs and routers, single-board computers, and on-board processors for deployment in safety-intensive applications [14] [15].

The rest of this paper is organized as follows. Section 2 provides a succinct discussion of the TMR scheme and briefly describes the majority voter functionality. Section 3 presents a number of standard-cell based majority voter designs including the existing and proposed ones. Section 4 compares the design metrics of 14 structurally different majority voters which are realized using a 32/28nm CMOS process. Lastly, Section 5 concludes this paper with a highlight of scope for further work.

## 2  TMR Scheme

TMR is a generic fault-tolerant design method that can be applied for combinational and/or sequential logic of a digital design. Also, different TMR architectures exist [17]. In a local TMR architecture, the sequential elements are alone triplicated and majority voting is performed at the outputs. In a distributed TMR architecture, both combinational and sequential logic elements are triplicated and their outputs are majority voted. In a global TMR architecture, combinational and sequential logic, and majority voters and buffers are all triplicated.

At the heart of any NMR scheme is a majority voter [18] [26], which, if fault-free, would confirm the correct operation of a majority of the function modules. Similarly, at the heart of any TMR scheme is a majority voter [26] which displays the correct functioning of at least two out of three (identical) versions of a function module by making a majority-based decision and conveying it through the output, as shown in Figure 1. Here, the term 'function module' may refer to any circuit or system that is triplicated to implement a TMR circuit/system. In Figure 1, function modules 2 and 3 are identical to function module 1. X, Y and Z represent the respective (and equivalent) outputs of function modules 1, 2 and 3 and also represent the inputs to the majority voter whose primary output is specified as V. The input and output labels (X, Y, Z, and V) shall be uniformly maintained throughout this paper for all the majority voter circuits. Assuming if any arbitrary function module becomes faulty/fails, the TMR implementation would still guarantee the correct operation on account of Boolean majority, which is established by the majority voter through the following logic equation. Notice that all the majority clauses are listed in equation (1), where product implies logical conjunction and sum implies logical disjunction.

$$V = XYZ + XY + YZ + XZ = XY + YZ + XZ \quad (1)$$

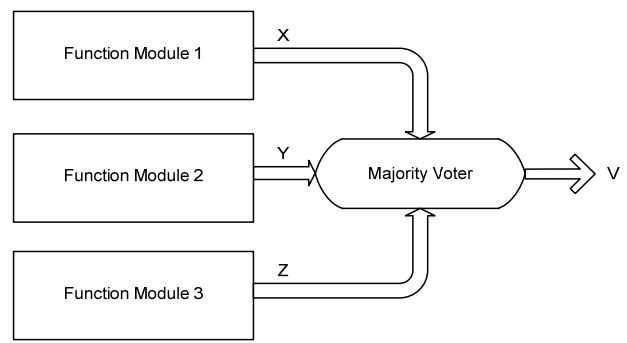

Fig. 1 Block diagram of the TMR scheme

## 3  Majority Voters

Provided only one function module becomes faulty/fails out of three identical function modules in a TMR implementation, the majority voter would mask the single function module fault/failure from being noticed by the external environment and manages to keep the entire system in correct operation. We shall now consider 14 structurally different majority voter designs.

### 3.1 Classical/Conventional Voters

The classical majority voter [6] [16], shown in Figure 2(a), consists of three 2-input AND gates in the first logic level and a 3-input OR gate in the second logic level, which synthesizes (1). This majority voter shall be referred by the acronym, AO_MV, for brevity. The acronym 'MV' implies 'Majority Voter', and shall be used in conjunction with the acronyms of different majority voters in this paper. Figure 2(b) portrays an alternative gate level representation of (1) which is easily derived by applying Boolean algebraic manipulations. This majority voter shall be labeled as NAND_MV. The NAND_MV was found to have the least SET sensitivity among 3 cell-based majority voter designs viz. NAND_MV, AO222_MV and BN_MV (which are described below), when subjected to a heavy ion irradiation campaign as described in [19].

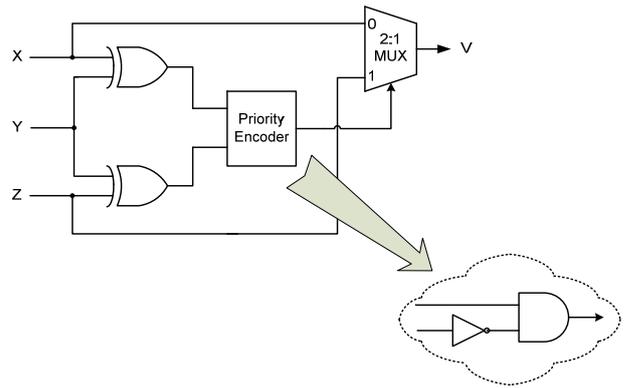

Fig. 3 Priority encoding based voter: KP_MV

### 3.3 Ban and Naviner Voter

Ban and Naviner [21] presented a majority voter shown in Figure 4(a), which shall henceforth be referred to as BN_MV. The BN_MV consists of just two gates – a 2-input XOR gate and a 2:1 MUX. Primary inputs X and Y of the majority voter are XORed and given as the select input for the 2:1 MUX. If the select input is 0, then input Y will be selected and its value will be forwarded to output V. However if the select input is 1, the majority voter input Z will be reflected as the output. An equivalent majority voter circuit employing a XNOR gate instead of the XOR gate, derived from the carry output logic of the XNOR-XNOR-based full adder [22] and the XNM-based full adder of [23], referred to as XNM_MV, is shown below in Figure 4(b).

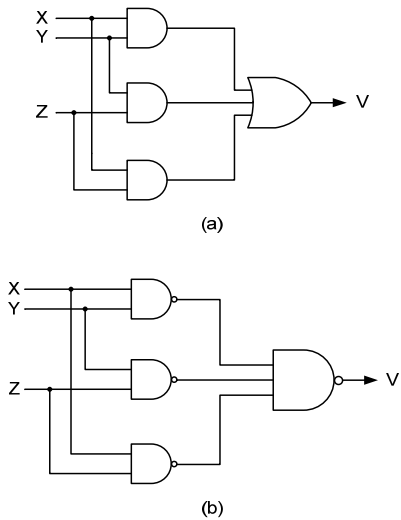

Fig. 2 Conventional voter circuits: (a) AO_MV and (b) NAND_MV

### 3.2 Kshirsagar and Patrikar Voter

The priority encoding based majority voter, proposed by Kshirsagar and Patrikar [20], henceforth identified as KP_MV is shown in Figure 3. Two 2-input XOR gates, a priority encoder (consisting of an inverter and a 2-input AND gate, as shown within the combinational cloud in dotted lines), and a 2:1 multiplexer (MUX) constitute the KP_MV. The KP_MV was shown to be more fault-tolerant than the conventional majority voter in [20].

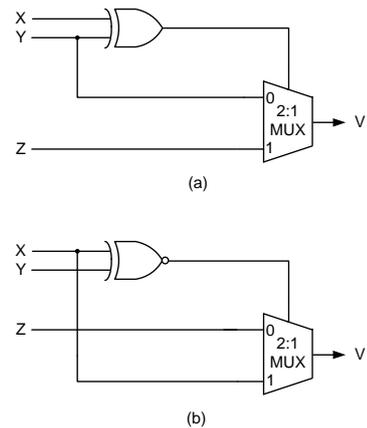

Fig. 4 (a) BN_MV; (b) XNM_MV

### 3.4 Simple and Complex Logic Gates Voters

This sub-section presents 4 majority voter designs, which are constructed using simple logic gates, or through a mix of simple and complex logic gates.

Figure 5(a) shows a majority voter design which comprises a 2-input XOR gate in the 1st logic level,

two 2-input AND gates in the 2$^{nd}$ logic level, and a 2-input OR gate in the 3$^{rd}$ logic level – this majority voter is identified by the acronym X2AO_MV. It may be noted that this majority voter corresponds to the typical carry-output logic of a full adder. The two 2-input AND gates in the 2$^{nd}$ logic level and the 2-input OR gate in the 3$^{rd}$ logic level, shown within the dotted ellipse in Figure 5(a), can be combined into a complex logic gate, namely the AO22 gate, identified as $G_1$ in Figures 5(b) and 5(c).

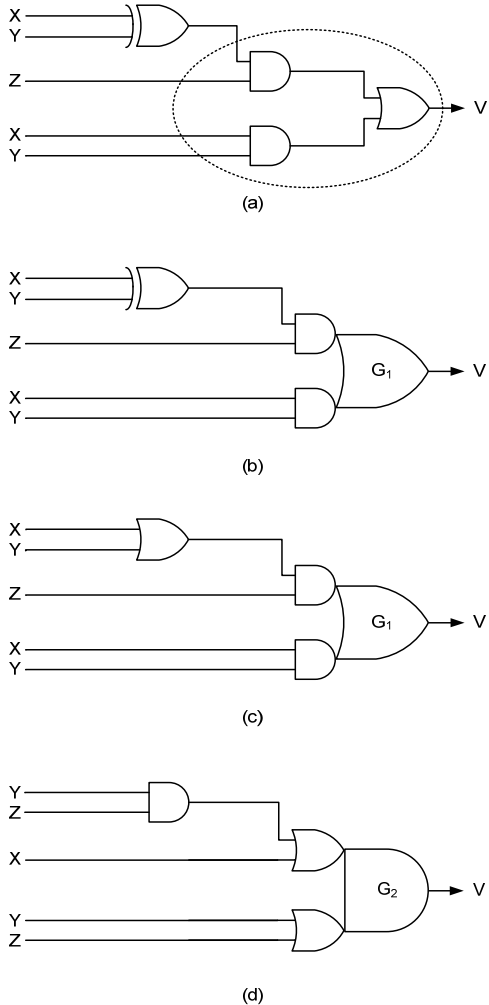

Fig. 5 Majority voters constructed using simple and/or complex logic gates: (a) X2AO_MV; (b) XAO22_MV; (c) OAO22_MV; (d) AOA22_MV

With A, B, C and D as inputs and E as the output, the AO22 gate implements the logic function E = AB + CD as a single entity. In static CMOS realization, two 2-input AND gates and a 2-input OR gate would require a total of 18 transistors, while the logically equivalent (single) complex gate AO22 would require just 10 transistors for physical realization, i.e. a 44% reduction in the device count is feasible, which tends to be advantageous from power, delay and area perspectives. In fact, the majority voter portrayed by Figure 5(b), which is labeled as XAO22_MV, resembles the carry-output logic of the full adder of [24].

Figures 5(a) and 5(b) synthesize (2), while Figure 5(c) showing the OAO22_MV synthesizes the factorized form [25] of (1), given as (3). It may be noted that (2) is in disjoint sum of products/sum of disjoint products form [32] – [34]. Figure 5(d), showing the AOA22_MV synthesizes (4). In Figure 5(d), the complex gate OA22 (marked as $G_2$) is used as a replacement for two 2-input OR gates and one 2-input AND gate, and it implements the function E = (A + B) (C + D) as a unitary gate, thus facilitating an approximate savings of 44% in the transistor count compared to a pure-discrete realization.

$$V = (X \oplus Y) Z + XY \qquad (2)$$

$$V = (X + Y) Z + XY \qquad (3)$$

$$V = (X + YZ) (Y + Z) \qquad (4)$$

### 3.5 Dual Complex Logic Gate Voters

This sub-section discusses two majority voters, which utilize the complex logic gates viz. OA21 and AO21. The OA21 gate, marked as $G_3$ and $G_6$ in Figure 6, realizes the functions (X + Y) Z and (Y + Z) K as single entities. The AO21 gate, highlighted as $G_4$ and $G_5$ in Figure 6, synthesizes the Boolean functions V = (XY + N) and K = (YZ + X) as unitary entities. The majority voter portrayed by Figure 6(a) shall be identified as OAAO_MV, while the majority voter shown in Figure 6(b) shall be referred to as AOOA_MV. The OAAO_MV and AOOA_MV synthesize equations (3) and (4).

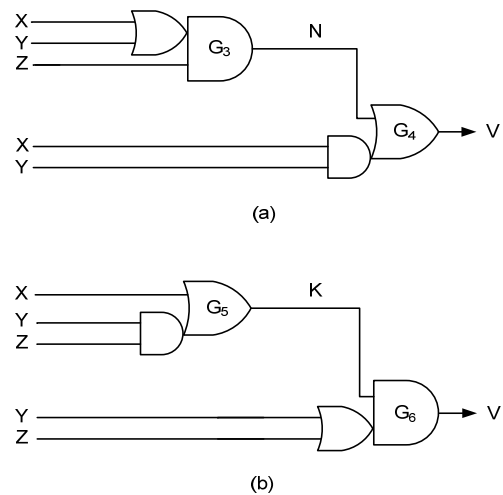

Fig. 6 Dual complex gate voters: (a) OAAO_MV; (b) AOOA_MV

### 3.6 Single Complex Logic Gate Voters

Thus far, majority voters designed using simple logic gates or using combinations of simplex and complex logic gates or utilizing just complex logic gates have been considered. In this sub-section, majority voter designs making use of only a single complex gate or a MUX (which is also a complex gate) are presented.

Three complex gate based majority voter designs are shown in Figure 7, sequentially labeled as AO222_MV, which uses the complex gate AO222 highlighted as $G_7$ in Figure 7(a); OA222_MV, which uses the complex gate OA222, highlighted as $G_8$ in Figure 7(b); and MUX41_MV, which employs a 4:1 MUX as shown in Figure 7(c).

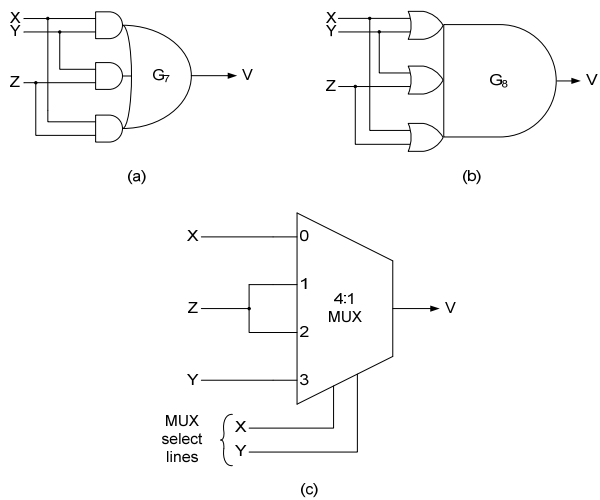

Fig. 7 Majority voters realized using a single complex gate: (a) AO222_MV; (b) OA222_MV; (c) MUX41_MV

The AO222_MV [26] resembles the carry-output logic of the minimum gates full adder shown in [27], and can be optimally realized in static CMOS style using 12 transistors based on equation (3) after logic factoring [36]. The AO222_MV was reported to be preferable for global TMR architectures in [19], where area occupancy may be a constraint. The OA222_MV constitutes another complex gate based design of the majority voter functionality, synthesizing (4), which also requires only 12 transistors for physical realization in static CMOS style after logic factoring [36]. The MUX41_MV is conceived on the basis of a traditional MUX-based implementation of logic functions and is technology mapped to the 4:1 MUX, made available as a complex gate in the digital cell library [28]. The MUX41_MV synthesizes (5) that exhibits logical equivalence with all the other majority voter equations mentioned thus far.

$$V = X(X'Y') + Z(X'Y) + Z(XY') + Y(XY) \quad (5)$$

## 4 Physical Realization and Results

A cell-based semi-custom implementation of the majority voter designs was considered. All the majority voters were described in accordance with the respective gate-level schematics shown, and their structural integrity was preserved during technology mapping based on the 32/28nm CMOS cell library [28]. This is to ensure a legitimate comparison of the design attributes of different majority voters after physical synthesis. The standard library cells [28] are inherently optimized for low power design. The power, delay and area results obtained for the different majority voters are given in Table 1.

Table 1. Average power dissipation, maximum propagation delay, area occupancy, and FOM of different majority voter designs

| Type of majority voter | Power (µW) | Delay (ns) | Area (µm²) | FOM |
| --- | --- | --- | --- | --- |
| AO_MV | 3.518 | 0.13 | 8.39 | 26.06 |
| NAND_MV | 1.564 | 0.10 | 6.35 | 100.69 |
| KP_MV | 6.286 | 0.30 | 15.25 | 3.48 |
| BN_MV | 3.488 | 0.22 | 7.62 | 17.10 |
| XNM_MV | 3.225 | 0.21 | 7.62 | 19.38 |
| X2AO_MV | 3.804 | 0.23 | 10.42 | 10.97 |
| XAO22_MV | 2.484 | 0.20 | 6.86 | 29.34 |
| OAO22_MV | 1.630 | 0.13 | 4.57 | 103.26 |
| AOA22_MV | 1.421 | 0.13 | 4.57 | 118.45 |
| OAAO_MV | 2.256 | 0.12 | 5.08 | 72.71 |
| AOOA_MV | 1.497 | 0.13 | 5.08 | 101.15 |
| AO222_MV | 1.207 | 0.12 | 3.30 | 209.22 |
| OA222_MV | 1.111 | 0.10 | 3.30 | 272.75 |
| MUX41_MV | 1.545 | 0.17 | 5.59 | 68.11 |

The design metrics evaluated correspond to a typical case PVT specification with recommended supply voltage of 1.05V and operating temperature of 25ºC. For total (average) power estimation, more than 1000 input vectors corresponding to a random sequencing of input patterns were identically supplied to the different majority voters at time intervals of 1ns (1GHz) through test benches, which represent the inputs coming in from the external environment. The value change dump (.vcd) files generated through the functional simulations were subsequently used for accurate average power estimation using Synopsys PrimeTime by invoking the time-based power analysis mode. The maximum propagation delay (i.e., critical path delay) and Silicon area were also estimated with suitable wire loads included automatically whilst performing the simulations. Minimum-sized discrete and complex gates (including 2:1 and 4:1 MUXes) of the cell library [28] were chosen uniformly for the different majority voters and their outputs were assigned with fanout-of-4 drive strength. To comprehensively comment on the design parameters of different majority voter designs, a figure-of-merit (FOM) is defined as the inverse of the product of power, delay, and area (i.e., $PDAP^{-1}$) [29] [30] [31] [37]. It has been shown in [29] [30] [31] [37] that FOM is an appropriate comprehensive parameter to quantify the physical attributes of a digital design. Since minimization of power, delay, and area is desirable, a lower PDAP value and thus a higher FOM value can be considered to be an indicator of an optimized design. The calculated FOM values, scaled up by a factor of 100, are also given in Table 1.

From Table 1, it may be evident that six majority voter designs viz. NAND_MV, OAO22_MV, AOA22_MV, AOOA_MV, AO222_MV, and OA222_MV possess above-average FOM values, while the other majority voter designs have below-average FOM figures, where the average FOM is calculated to be 82.33. It may be interesting to note that among the eight majority voter designs viz. XNM_MV, X2AO_MV, XAO22_MV, OAO22_MV, AOA22_MV, OAAO_MV, AOOA_MV, and OA222_MV, four of them feature above-average FOM values. In general, majority voter circuits constructed predominantly using discrete gates tend to have less FOM values, while majority voters constructed using complex gates possess high FOM values.

Among the different majority voter designs, the OA222_MV, which is one of the proposed majority voter designs reportedly has the highest FOM of 272.75, thanks to pre-logic factoring [36] followed by physical synthesis, with the AO222_MV having the second-best FOM of 209.22 – in comparison with the latter, the former features enhanced FOM of 30.4%. In comparison with the traditional and other existing majority voter designs viz. NAND_MV, KP_MV and BN_MV, the OA222_MV reports significant improvements in FOM by 2.7×, 78.4×, and 16× respectively. The KP_MV has the lowest FOM of 3.48 – this is a direct consequence of it requiring more number of gates for physical realization consequently resulting in more area occupancy and power dissipation and also more propagation delay due to more number of logic levels, thus adversely affecting its FOM.

## 5 Conclusions and Further Work

In a recent study [19], SET tolerance of six 65nm CMOS majority voters was evaluated when subjected to heavy ion irradiation experiments, and it was found that the NAND_MV is more SET-tolerant than any other analog or digital voter considered. It was also reported in [19] [26] that the AO222_MV has optimized design metrics and is therefore suitable for global TMR architectures. In contrast, this work has shown that the OA222_MV is optimal in terms of design metrics amongst 14 structurally different voter designs including the AO222_MV. This paper has presented 14 majority voter designs including the existing and proposed ones and has made a fair comparison of their design attributes based on physical implementation using a high-end 32/28nm CMOS process. Among the various majority voter designs presented, the OA222_MV reportedly yields the best FOM. This confirms our earlier observation [35] that in some cases the product of sums (POS) form of a Boolean function could lead to a lower power design than the sum of products (SOP) form. The OA222_MV, which corresponds to the POS form, is found to be a lower power design having better FOM than the AO222_MV which is based on the SOP form of the majority voter functionality.

Subsequently, this paper aims to motivate further research into the practical measurement of SET tolerance of all the 14 majority voter designs discussed in this work based on a similar heavy ion irradiation campaign as carried out in [19], since [19] has considered only a few majority voter designs. This could be useful for determining the best candidate majority voter(s) with respect to FOM and/or radiation (SET) tolerance for local, distributed, and global TMR architectures for deployment in ASIC-based mission/safety-critical circuit and system designs. Also, tradeoffs between FOM and radiation tolerance of the majority voters can be analyzed. Further, design of majority voters for the TMR architecture can be explored based on the gate-diffusion-input technique [38] – [40], and their FOM and radiation tolerance can be evaluated vis-à-vis static CMOS majority voters.